\documentclass[aps,prl,preprint,groupedaddress]{revtex4-2}
\usepackage{hyperref}
\usepackage{graphicx}

\begin{document}

\title{Time-Resolved Phonon Dispersions of Thermally Excited Graphite at the Nanoscale}
\author{Mairi McCauley}
\author{Christoph T. Koch}
\author{Benedikt Haas}

\affiliation{Department of Physics and Center for the Science of Materials Berlin, Humboldt-Universit\"at zu Berlin, Berlin, Germany}

\date{\today}

\begin{abstract}

    Lattice vibrations govern the propagation of heat in materials, and investigating their dynamics through time-resolved phonon dispersion measurements would provide insight into these excitations and their interactions. However, time-resolved measurements that directly resolve the phonon dispersion at the nanoscale have not been previously demonstrated.
    Momentum-resolved vibrational electron energy loss spectroscopy (EELS) provides access to phonon energies and momenta with nanoscale spatial resolution. Here, we combine momentum-resolved vibrational EELS with pulsed laser heating and synchronised gated electron detection to observe the phonon dispersion of graphite during controlled heating and cooling on the microsecond timescale. 
    Recording both the gain and loss spectra at defined times relative to laser excitation, we determine the temperature at each time delay and follow changes to the phonon branches as a function of temperature. The momentum-resolved spectra reveal a decrease in energy of the optical phonon branch with increasing temperature accompanied by an increase in peak width. These changes are consistent with temperature-dependent phonon renormalisation and scattering associated with anharmonic interactions. The approach is compatible with detector gating on timescales down to 50 ns, providing a route towards extending these measurements to faster dynamics. This work demonstrates time-resolved phonon dispersion spectroscopy using vibrational EELS and provides a route towards probing temperature-dependent phonon dynamics at the nanoscale with simultaneous energy and momentum resolution. 

\end{abstract}

\maketitle
\newpage

\section{Introduction}
Monochromated electron energy-loss spectroscopy (EELS) has established vibrational spectroscopy in the electron microscope, providing access to phonon excitations at the nanometre and atomic scale \cite{Haas2024}. The development of monochromated and aberration corrected scanning transmission electron microscopy (STEM) has enabled energy resolutions down to a few meV, allowing for low energy excitations to be directly resolved \cite{Krivanek2014}. 
Vibrational EELS has subsequently been used to measure phonon excitations with atomic spatial localisation \cite{Hage2020} and to investigate the influence of interfaces \cite{Hoglund2022,Gadre2022}, defects \cite{Jiang2024} and grain boundaries on the phonon properties \cite{Haas2023}.

The temperature dependence of phonon dispersions provides information about the thermal transport and scattering dynamics of a specimen. As the temperature increases, obviously the occupation of higher energy phonon states increases according to Bose-Einstein statistics, but phonon modes and their lifetimes can also change due to anharmonic interactions such as phonon-phonon scattering. For example, temperature-dependent Raman measurements of graphite have demonstrated a decrease in energy of a Raman-active vibrational mode with increasing temperature, alongside a change in linewidth arising from anharmonic interactions and phonon-electron coupling \cite{Giura2012,Montagnac2013}.
Probing the temporal evolution of phonon dispersions during heating and cooling enables a deeper understanding of the processes governing phonon scattering.

Vibrational EELS also provides access to the thermal population of phonons through electron energy gain spectroscopy (EEGS). During this process, an electron transmitted through the specimen can gain energy from annihilation of a thermally populated excitation or lose energy by exciting this state. The ratio of the energy-gain and energy-loss intensities is related to the phonon occupation by the principle of detailed balance (PDB) and can therefore be used to determine the local specimen temperature \cite{Lagos2018,Idrobo2018}. This approach has recently been used to measure the interfacial thermal resistance between two layers \cite{Liu2025}, investigate the temporal heat dissipation \cite{Castioni2025} and extract nanoscale thermal properties \cite{McCauley2026}.

Momentum-resolved vibrational EELS provides the capability to resolve not only in energy, corresponding to a projected phonon density of states, but to obtain phonon dispersion curves showing individual phonon branches and potentially their lifetimes. Phonon dispersions in graphite and similar materials have been measured using an off-axis detection scheme and sequentially acquiring spectra for different momenta \cite{Senga2019, Hage2018}. Subsequent approaches utilised a rectangular aperture to simultaneously acquire spectra for a continuous momentum range, recording the full phonon dispersion between Bragg spots along a specific momentum direction in a single acquisition \cite{Lovejoy2019,Lovejoy2020}. This approach was later applied to study phonon dispersion at an interface \cite{Qi2021} and investigate magnons \cite{Kepaptsoglou2025}. The method has been extended to cover a two-dimensional momentum range by displacing the diffraction pattern relative to the aperture \cite{Haas2023_MM}. More recently, time- and momentum-resolved EELS of electronic excitations at about 2 eV energy resolution has been demonstrated in graphite to investigate ultrafast electronic and plasmon dynamics \cite{Barantani2025}. These developments demonstrate the potential of combining temporal and momentum resolution in electron scattering.

Here, we combine stroboscopic laser heating with momentum-resolved vibrational EELS and time-resolved detection to follow the evolution of the phonon spectrum of graphite during a heating and cooling cycle. By synchronising the optical excitation with gated electron detection, momentum-resolved vibrational EELS spectra are acquired at defined times during specimen heating and cooling. The temperature at each time is determined by fitting the phonon population, allowing us to follow the evolution of phonon branches as a function of temperature and momentum. Due to the limited time-resolution of the setup (50 ns) compared to phonon scattering times (typically fs), the phonons exhibit thermalised occupation. The measured optical phonon energy decreases with increasing temperature, whilst the peak width increases. These observations are consistent with temperature-dependent phonon renormalisation and scattering associated with anharmonic interactions \cite{Giura2012,Han2022}. More generally, these measurements demonstrate the first time-resolved phonon dispersion measurements and show the potential of the method for studying phonon dynamics and non-equilibrium populations. 

\begin{figure}[ht]
    \centering
    \includegraphics[width=\linewidth]{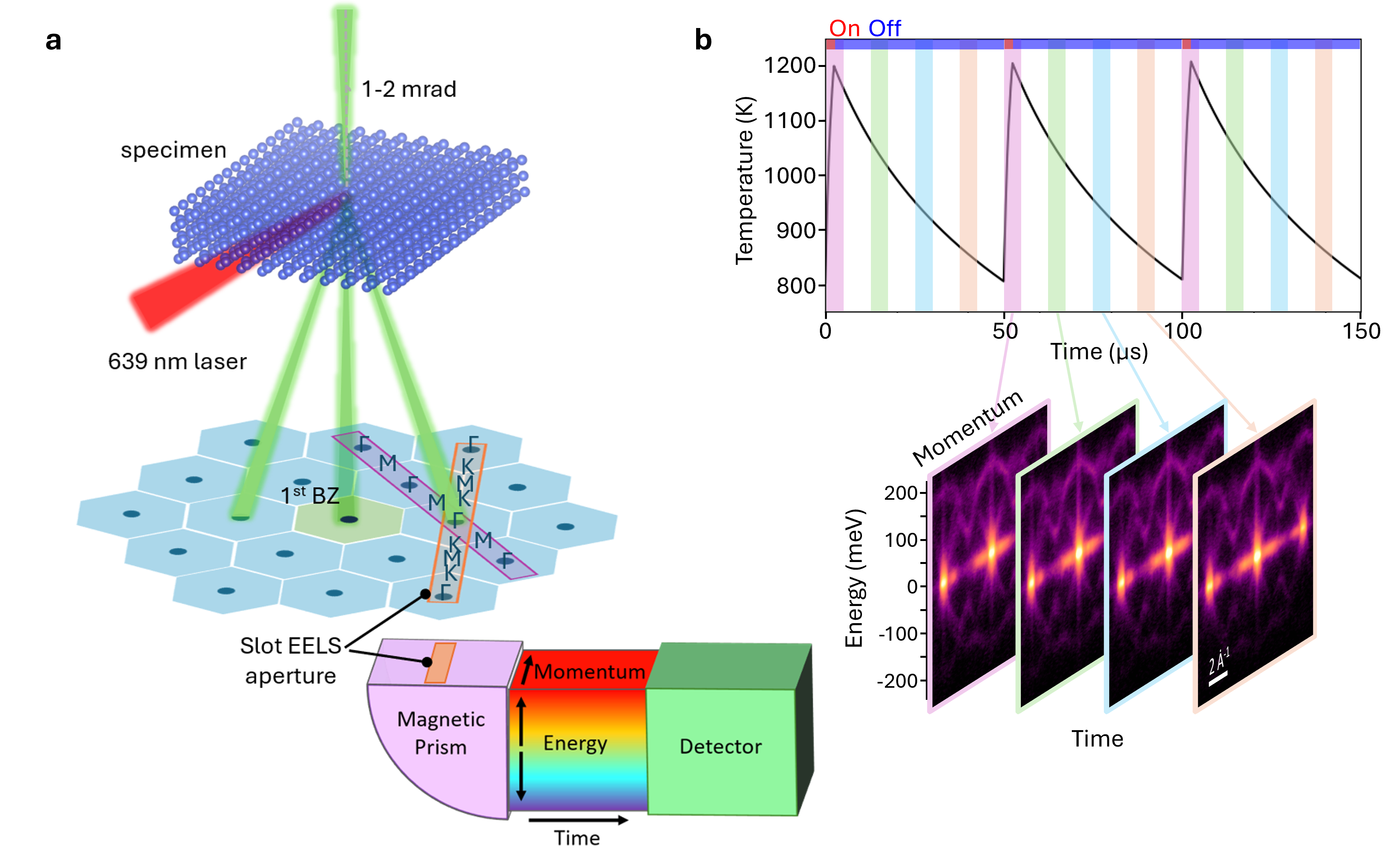}
    \caption{a) Schematic of the momentum resolved EELS setup where an electron beam with a small convergence angle is transmitted through the sample and a slot EELS entrance aperture is positioned along a high symmetry axis in the diffraction plane. The reciprocal space is schematically sketched for graphite. Excitation of the sample with a 639 nm wavelength laser is synchronised with gated detection to obtain temporal information. b) Synchronised laser excitation and electron detection scheme indicating heating and cooling over three laser on/off cycles. Gating windows used to acquire time-resolved dispersions are overlaid.}
    \label{fig:exp_concept}
\end{figure}

\section{Methods}
Two samples were used for the measurements. The main sample was prepared by dispersing graphene nano-platelets (Graphene Nanoplatelets 500, ProGraphite GmbH) in isopropanol and depositing the dispersion onto a carbon Quantifoil grid. Additional measurements were performed on a graphite flakes exfoliated from a crystal and supported by a lacey carbon grid.

EELS measurements were performed using a Nion HERMES STEM operated at 60 kV. The microscope was monochromated to a zero-loss peak (ZLP) width of 11 meV measured at $\Gamma$ in the static momentum-resolved datasets and 11 meV in the time-resolved datasets, with a beam current of approximately 2 pA. A convergence semi-angle of 2.0 mrad was used for measurements on the Quantifoil sample and a convergence semi-angle of 1.6 mrad used for measurements on the lacey C sample. A slot aperture in the diffraction plane was used to select the momentum transfers as illustrated in Fig. \ref{fig:exp_concept}a. The momentum scale was calibrated from the positions of the diffraction spots, and the corresponding momentum transfer was determined relative to the $\Gamma$ point. 

Time-resolved phonon dispersion maps were acquired using the same momentum-resolved EELS configuration as for static measurements. The laser injection system, laser alignment and general time-resolved measurement scheme are described in detail in a previous publication \cite{McCauley2026}. The specimen was excited using a laser diode of 639 nm wavelength pulsed with a duration of 2.5 $\mu$s and peak power density of $\approx$86.6 MW/m$^2$ at a repetition rate of 20 kHz, corresponding to a 50 $\mu$s repetition period and 5\% duty cycle. The Dectris ELA detector was gated at defined time delays relative to the laser pulses. Gated windows of 5 $\mu$s were used and summed over 2000 triggers for a single frame (10 ms exposure, 100 ms total time). At each time delay 600 frames were acquired, aligned and summed together to achieve sufficient signal to noise (60 s exposure time, 600 s total time). The heating and cooling cycle with the time windows used for EELS measurements is shown in Fig. \ref{fig:exp_concept}b. 

The resulting two-dimensional momentum-resolved spectra were aligned and distortion-corrected by shifting the energy axis so that the ZLP was positioned at 0 meV for each momentum transfer. The displayed spectra were otherwise unprocessed and are shown on a logarithmic intensity scale. Representative time-resolved momentum-resolved spectra are shown in Fig. \ref{fig:exp_concept}b. 

Temperature measurements were performed by fitting the relative intensities on the energy-gain and energy-loss sides of the EELS spectrum to the Bose-Einstein statistics. For thermally populated vibrational modes, the ratio of the loss and gain intensities is related to the specimen temperature according to $\exp\left(\frac{E}{k_{\mathrm{B}}T}\right)$, where $E$ is the vibrational excitation energy, $k_{\mathrm{B}}$ is the Boltzmann constant and $T$ is the specimen temperature. The ratio of the loss to gain intensities, after subtraction of a fitted background function, was fitted over the phonon energy range and the temperature calculated. Uncertainty in measured temperature was determined from the least-square errors of the linear fit of the ratio against energy. The background was modelled as an asymmetric Pearson VII function fitted to the ZLP. 

For quantitative analysis, EELS spectra were extracted at selected momentum positions by summing four detector pixels, corresponding to a momentum interval of 0.26 \AA$^{-1}$ unless otherwise stated. The spectra were fitted in Python using Hyperspy \cite{hyperspy}. A power-law background and two Voigt functions were used to describe the acoustic and optical phonon contributions respectively. At the $\Gamma$ point where the acoustic phonon energy approaches zero, only the optical phonon contributions were fitted. The phonon energies, peak widths and uncertainties were determined from the least-squares fitting procedure. 

\section{Results}

\subsection{Static Measurements}

\begin{figure}[ht]
    \centering
    \includegraphics[width=\linewidth]{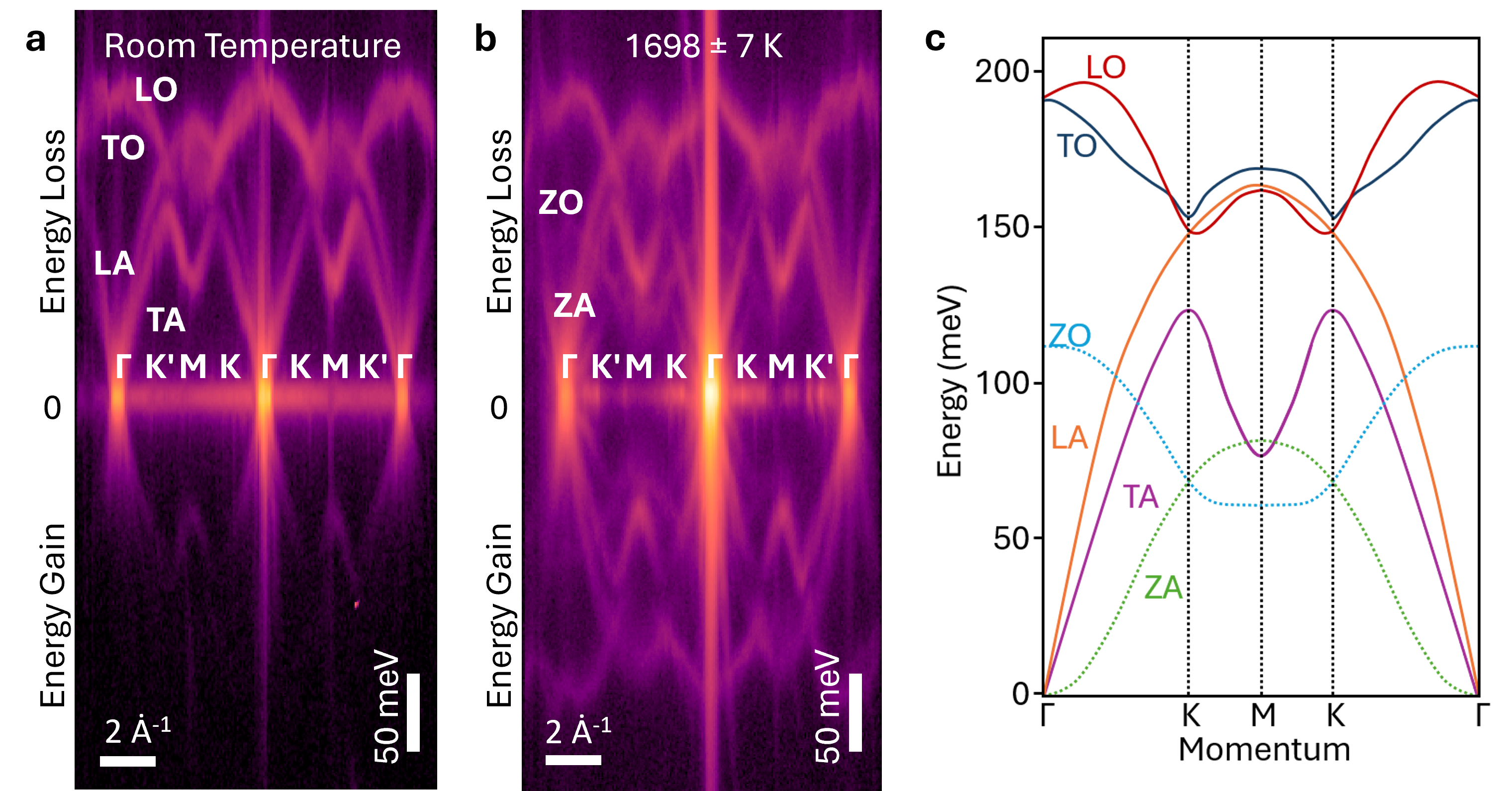}
    \caption{Phonon dispersion of graphite (lacey C supported sample) along $\Gamma$KM. a) Phonon dispersion acquired at room temperature. b) Phonon dispersion during laser heating where the temperature was measured as 1698$\pm$7 K. c) Theoretical dispersion for the $\Gamma$KM momentum direction \cite{Senga2019}.}
    \label{fig:static_1}
\end{figure}

Momentum-resolved EELS data obtained at room temperature from a graphite crystal (lacey C sample), shown in Fig. \ref{fig:static_1}a reveal the phonon dispersion along the $\Gamma$KM momentum direction. To perform these measurements, the diffraction pattern was displaced to avoid the intense central beam as illustrated in Fig. \ref{fig:exp_concept}a. In both datasets four distinct phonon branches are observed with two acoustic modes dispersing from 0 meV at $\Gamma$ and two optical modes at approximately 194 meV at $\Gamma$. The experimental results are in agreement with the dispersions from theory, shown in Fig. \ref{fig:static_1}c, where 3 non-degenerate acoustic (TA,LA,ZA) and 3 non-degenerate optical (TO,LO,ZO) phonon branches are expected \cite{Senga2019}. In the given sample orientation with the beam perpendicular to the planes, out-of-plane modes (ZO and ZA) cannot be observed in EELS because the electron beam is insensitive to vibrations parallel to its direction of travel. The transverse optical TO and longitudinal optical modes LO are expected to disperse between approximately 197 meV at $\Gamma$ to 163/153 meV at $K$ and 177/170 meV at $M$ respectively \cite{Senga2019}, whilst the transverse acoustic (TA) and longitudinal acoustic (LA) modes disperse from 0 meV at the $\Gamma$ point to approximately 124/153 meV at the $K$ point and 78/167 meV at $M$ respectively. 

\begin{figure}[ht]
    \centering
    \includegraphics[width=\linewidth]{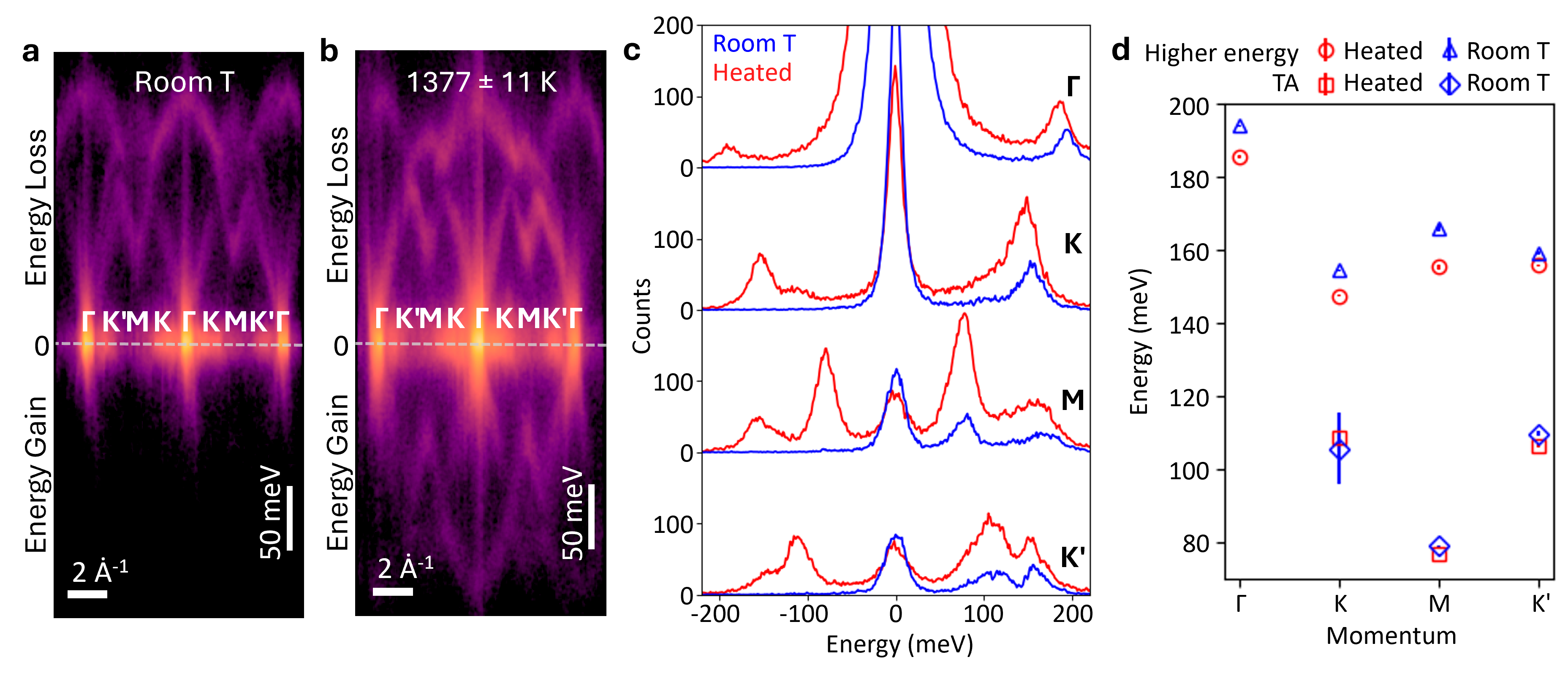}
    \caption{Phonon dispersion of a graphite flake (Quantifoil sample) at a) room temperature and b) 1377$\pm$11 K from laser heating. c) Spectra extracted from the room temperature and 1377$\pm$11 K measurements in a,b at the $\Gamma$, K, M and K' points indicated in the dispersion plots. d) Fitted energy of the higher energy (TO/LO at $\Gamma$, TO/LO/LA at other positions) and lower energy (TA) phonon peaks.}
    \label{fig:static_2}
\end{figure}

At the 0 meV line the intensity varies and is not solely localised at the $\Gamma$ points as would be the case for diffraction from a single crystal. In most momentum-resolved EELS datasets, a continuous line of faint intensity lies along the 0 meV line between $\Gamma$ points. Momentum-resolved spectra acquired on this and similar samples prior to heating, such as the one shown in Fig. \ref{fig:static_1}a, exhibit this intensity at 0 meV. Upon heating and subsequent cooling, this continuous intensity band transforms into discrete intensity points and regions with reduced or no intensity. This behaviour supports the hypothesis that this band originates from amorphous surface contaminants on the specimen, which distribute the elastically scattered beam across momenta away from the Bragg positions of the crystal. The diffuse diffraction rings of an amorphous contamination layer are additionally smeared by the slot positioned as a secant of these rings (due to avoiding the direct beam). Heating can cause this amorphous layer to become polycrystalline, as demonstrated by the authors earlier \cite{McCauley2026}. This leads to regions of enhanced and reduced intensity between crystal $\Gamma$ points in the spectrum that persist after cooling down. The room temperature data presented later in Fig. \ref{fig:static_2}a was acquired after the specimen had been previously heated, explaining the uneven intensity between $\Gamma$ points.

During heating with a continuous wave laser of power density $\approx$16.6 MW/m$^2$ the spectrum shown in Fig. \ref{fig:static_1}b was acquired. The most obvious difference between the momentum-resolved spectra at room temperature and increased temperature shown in Fig. \ref{fig:static_1}a and \ref{fig:static_1}b is the significant increase in signal intensity, most notable in the gain. This increase reflects the Bose-Einstein occupation of vibrational modes. The characteristic thermal energy scale, $k_BT$, increases from approximately 25 meV at room temperature to approximately 120 meV at 1400 K, approaching the phonon energies in graphite investigated here. The equilibrium phonon occupation, given by $n(E,T)=1/(exp(E/k_BT)-1)$, increases with temperature and the gain- and loss-intensities are proportional to $n$ and $n+1$ respectively. The specimen temperature of 1698$\pm$7 K was determined from the gain- and loss-intensities using the PDB as described in Methods.
The flake was supported only by strings of carbon, which, upon heating, thermally expanded introducing a large tilt. The tilt was mainly perpendicular to the long slot axis as can be seen from the changed overall diffraction intensity and the conservation of symmetry on either side of the middle Bragg spot. Due to this tilt the ZO and ZA phonon branches can now be observed.

Phonon dispersions acquired on a different graphite sample, supported on a carbon Quantifoil grid for stability, at room temperature and with $\approx$12.8 MW/m$^2$ laser excitation are shown in Fig. \ref{fig:static_2}a,b. As expected the gain intensity increases substantially under laser heating and the specimen temperature was measured as 1377$\pm$11 K. The lower specimen temperature compared to the previous sample despite higher laser power density was due to the greater thermal conduction of the supporting carbon and closer proximity to a copper grid bar acting as a heat sink. The laser spot was partially incident on the grid bar reducing the heat absorption by the carbon support. For the LO/TO modes at approximately 190 meV, the increase in temperature from 300 K to 1377 K increases the occupation by a factor of almost 400.

Although not all individual phonon branches are fully resolved at all momenta in the spectrum in Fig. \ref{fig:static_2}b, the optical and acoustic modes could be fitted at different momenta and generally agree with the expected theoretical result. At positions where phonon branches are degenerate or very close in energy, the fitted peak contains contribution from multiple phonon branches. For example, the TO and LO modes contribute to a single peak at $\Gamma$ and at $K$ and $M$ points the TO, LO and LA modes all contribute to the higher energy peak.

Beyond the expected changes in phonon occupation, heating induces additional, more subtle modifications to the graphite phonon spectrum. A comparison between room temperature and heated spectra extracted from phonon dispersion plots at different high symmetry points with momentum width of 0.26 \AA$^{-1}$ is shown in Fig. \ref{fig:static_2}c. At the $\Gamma$ point a redshift of the optical mode from 193.7$\pm$0.2 meV at room temperature to 185.2$\pm$0.3 meV at 1377$\pm$11 K is observed with similar redshifts observed to this mode at other momentum positions as shown in Fig.\ref{fig:static_2}d. 
In addition, this peak is observed to broaden at elevated temperatures. The observed energy shift to the optical mode is significant, however smaller shifts to the acoustic modes at elevated temperature potentially observable at different momenta mainly fall within the measurement uncertainty.

\subsection{Time-Resolved Measurements}
Laser induced heating may occur directly by excitation of vibrational modes and in addition, or exclusively by, the excitation of electrons and subsequent electron-phonon scattering. The system gradually thermalises through electron-electron, electron-phonon and phonon-phonon scattering on femtosecond to picosecond timescales \cite{Yang2017}. As these processes are much faster than the timescales accessible here, the measurements probe subsequent diffusive heat transport and the phonon population can be considered approximately thermalised during the measurements. 

\begin{figure}[ht]
    \centering
    \includegraphics[width=\linewidth]{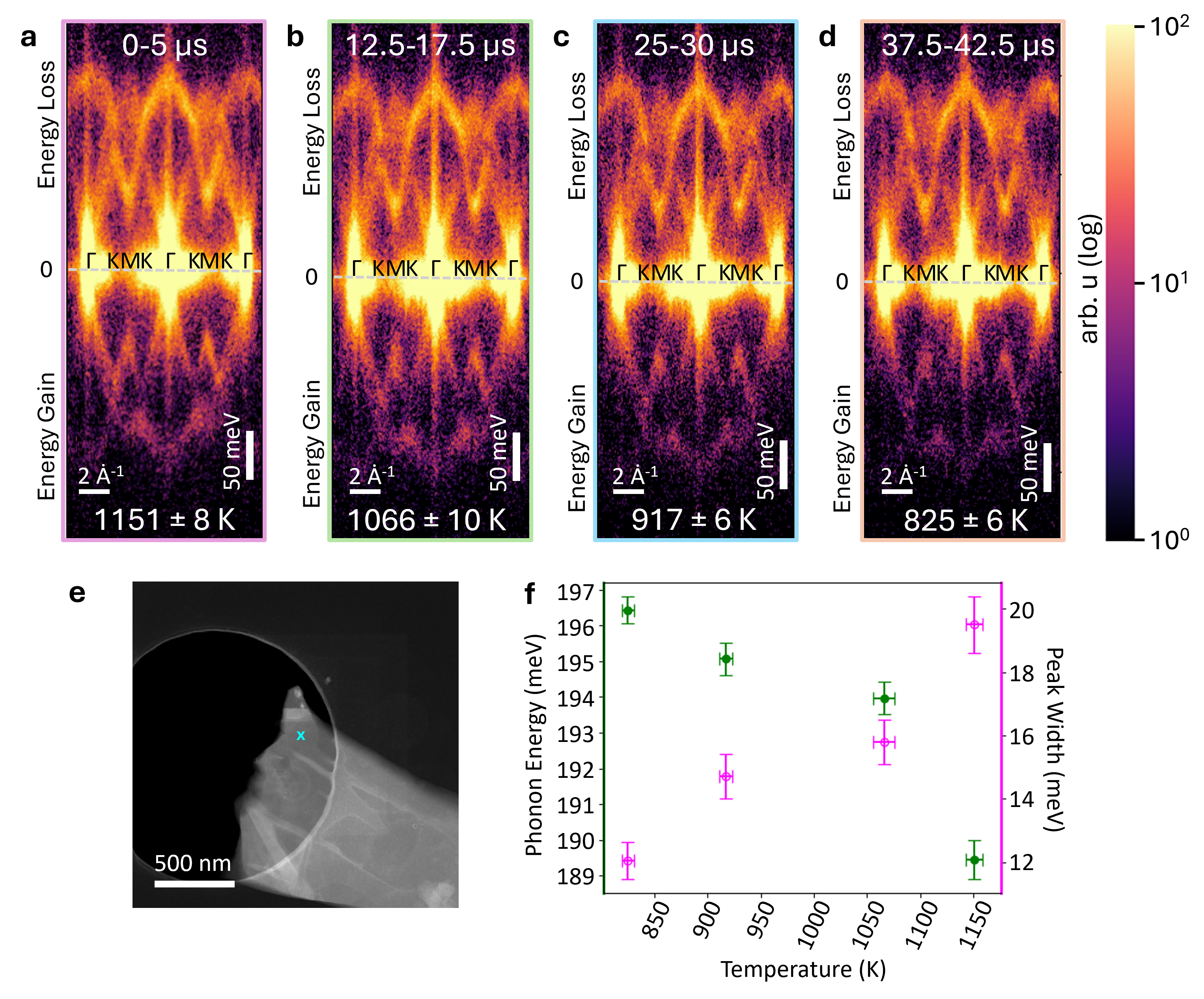}
    \caption{Time-resolved phonon dispersions from vibrational EELS. a-d) Time-resolved graphite phonon dispersion along $\Gamma$KM at different time delays relative to a 2.5 $\mu$s laser pulse. e) High-angle annular dark field (HAADF) image of the measured graphite flake supported on a carbon Quantifoil grid with the measurement position indicated by a blue cross. f) Fitted energy and width of the combined TO and LO phonon modes at $\Gamma$ against temperature.} 
    \label{fig:TR_results}
\end{figure}

Time-resolved phonon dispersion spectra, acquired as outlined in Methods, are shown in Fig. \ref{fig:TR_results}a-d at time delays relative to the beginning of a 2 $\mu$s laser pulse. All spectra were acquired from the same position on the 78 nm thick graphite crystal indicated in Fig. \ref{fig:TR_results}e. Specimen temperature, extracted by summing over all momenta in each spectrum and performing PDB analysis, was measured as 1151$\pm$8 K, 1066$\pm$10 K, 917$\pm$6 K and 825$\pm$6 K, decreasing with time. Additionally temperature measurements were made across the acoustic and optical phonon branches at distinct momenta to compare. The temperatures extracted from each phonon branch at each momenta were consistent with the summed temperature within the measurement uncertainty. This confirms that the system has reached a thermal equilibrium where the occupation of phonon modes obeys Bose-Einstein statistics. 

In addition to changes in the relative intensity of phonon branches in accordance with Bose-Einstein statistics, the energy of the high energy peak increases over time as the temperature decreases, as shown in Fig. \ref{fig:TR_results}f. An energy shift is observed in fitted spectra not only at the $\Gamma$ point in Fig. \ref{fig:TR_results}f but also at each momentum position. In addition to the energy shift, the width of the peak decreases with decreasing temperature. To confirm that this trend was not due to measurement artefacts or differences in alignment, room temperature and laser heated spectra of the lacey C supported sample shown in Fig. \ref{fig:static_1}a,b were fitted and also showed decreased energy and increased peak width at elevated temperatures. This trend was also observed in additional time-resolved measurements acquired on a different sample along the $\Gamma$M momentum axis (see Supplementary Information), Fig. \ref{fig:S2}.

\section{Discussion}

A systematic redshift in phonon peak energy was observed upon increasing temperature by laser excitation in both the static and time-resolved measurements. The energy shift was observed to be reversible upon cooling and heating as evidenced by the stroboscopic time-resolved results. Redshifts of phonon energies at elevated temperatures have been observed previously with vibrational EELS in SiC, where phonon energy was observed to redshift by 5 meV upon heating by approximately 1000 K \cite{Yan2019}. The magnitude of this redshift could not be explained by thermal lattice expansion alone and different orders of phonon-phonon scattering were found to play a major role.

Temperature-dependent energy shifts of vibrational modes in graphite have been observed using Raman and infrared spectroscopy. These optical techniques only probe the spectrum at $\Gamma$ where the optical phonon modes are termed according to their symmetry as E$_{2g}$ (Raman-active) and E$_{1u}$ (infrared-active). Experimental measurements have shown the redshift of these two modes by differing amounts as temperature is increased \cite{Giura2012,Montagnac2013} accompanied by changes to their linewidths. The infrared-active mode has been observed initially at higher energy than the Raman-active mode before decreasing in energy upon increasing temperature. This mode was observed to have a steeper decrease in energy with temperature than the Raman active mode. In EELS we can probe both Raman and infrared-active modes simultaneously, however, in graphite the infrared-active signal is expected to be very weak and therefore the Raman-active mode should dominate at $\Gamma$. Accompanying calculations find that the redshift can be explained by three contributing factors; three-phonon interactions, four-phonon interactions and thermal lattice expansion \cite{Bonini2007,Liu2019}. 

The phonon-phonon interactions are expected to increase substantially at elevated temperature due to the increased thermal occupation of phonon modes which increases the probability of scattering. The softening of the optical phonon mode in Fig. \ref{fig:static_2}d and the energy shifts observed in Fig. \ref{fig:TR_results}f during specimen cooling can be explained by the phonon anharmonicity discussed above. 

Phonon softening is not expected to be uniform across momentum and molecular dynamics simulations have indicated a momentum dependence on the magnitude of phonon energy shifts in graphene \cite{Koukaras2015}. It was found that optical phonon branches show the strongest temperature dependence and the magnitude of softening of acoustic branches increases near high symmetry $K$ and $M$ points. The momentum-dependence of phonon softening is reflected in the results presented in Fig. \ref{fig:static_2}d, where the magnitude of energy shift of the optical mode differs slightly at different momenta. In addition, the smaller magnitude of acoustic mode energy shift and differences across momentum is in agreement with this theory.

Accompanying the phonon softening in all measurements is the broadening of the peak with increasing temperature. Although the experimental energy resolution does not resolve the individual phonon branches at all momenta in the results presented here, the broadening trend is observed in different datasets and shows a consistent increase at higher temperatures. Therefore even though the fitted peaks contain contributions of multiple phonons, the temperature-dependent broadening can be further analysed. 

Optical experiments indicate differing temperature-dependent energy shifts of the observed Raman and infrared-active modes at $\Gamma$ \cite{Giura2012}. As both of these modes are excitable in EELS, observed peak broadening could be due to the increasing separation between these modes at increased temperature which would appear as peak broadening as the modes were not individually resolved. The initial energy separation of these modes at room temperature in literature is 1 meV which reduces to 0 at around 750 K and was theorised to increase as the peak energies diverge at higher temperature \cite{Giura2012}. The peak width at $\Gamma$ in the time-resolved data shown in Fig. \ref{fig:TR_results}f increases by around 7 meV could be explained by an increase in the relative separation of these modes. Although the TO and LO branches are degenerate at $\Gamma$, the finite momentum resolution of the measurements means that some modes with non-zero momentum transfer can also contribute to the peak and may show temperature dependence of their relative energies.

Another explanation could be the temperature dependence of the phonon linewidth. At higher temperatures, increased phonon scattering occurs shortening the phonon lifetime and hence broadening the linewidth. Raman spectra have observed a room temperature linewidth of around 1.4 meV which slightly decreases to 1.2 meV at about 600 K before slightly increasing again for higher temperatures  \cite{Liu2019}. The whole variation of less than 0.2 meV is well below our measurement uncertainties and therefore does not play a role for our analysis.
Similarly, the linewidth of the infrared-active mode has been measured to increase by a fraction of a meV over hundreds of Kelvin temperature increase \cite{Giura2012}. In graphene a similar observation has been made, with the increase Raman active mode linewidth attributed to the influence of higher order phonon-phonon interactions which increases at higher temperatures \cite{Han2022}.

The linewidth of phonon branches is strongly momentum-dependent and has been shown to be highly non-uniform in graphite and graphene calculations \cite{Paulatto2013}. The calculations showed abrupt changes to the linewidths with momentum due to allowed decay channels becoming available at certain momenta. Extraction of the temperature dependence of the linewidths of phonons not observed at $\Gamma$ such as acoustic modes, therefore requires momentum-resolution. Since the allowed decay channels depend on momentum, the increase in thermal occupation, increase in scattering and thus linewidth broadening should also have some momentum dependence. In the measurements presented here, the peak width of different phonon branches at different momenta are not the same, however, broadening with increased temperature is observed for different modes.

\section{Conclusion}

In conclusion, we have demonstrated time-resolved phonon dispersion measurements. By combining synchronised pulsed laser excitation with gated momentum-resolved EELS detection, phonon dispersions were acquired at defined times during a heating and cooling cycle of a graphite flake.

The momentum-resolved spectra reveal a redshift with increasing temperature, accompanied by increased phonon peak widths. These observations are consistent with temperature-dependent phonon softening and increased phonon scattering, and thus potentially reduced lifetimes, due to anharmonic phonon interactions. 

These measurements demonstrate the potential of time-resolved phonon dispersion spectra to study the evolution of phonon properties of transient states with nanoscale spatial resolution. Although the temporal resolution is currently limited to about 50 ns, the combination of this with nanoscale spatial resolution, meV energy resolution and momentum-resolution should enable the study of non-equilibrium systems. This provides a means of investigating the effect of thermal gradients and non-uniform heating on the properties of different phonon modes. 

Combining 50 ns temporal resolution with nanometre spatial resolution for phonon dispersions opens a route to probing how thermal history is encoded in local lattice dynamics. Promising targets include the relaxation of quenched charge-density-wave domains and the emergence of crystalline phonon branches during phase-change switching. Spatially resolved measurements across evolving domains and phase boundaries could establish how metastable order reshapes the vibrational spectrum, and how that spectrum evolves as structural memory is erased. Such measurements would access the pathways connecting thermally selected states, beyond the stationary spectra available at fixed temperatures.

\section{Acknowledgements}

The authors thank Johannes M\"{u}ller for providing the graphite sample on lacey carbon. This project was funded by the Deutsche Forschungsgemeinschaft (DFG, German Research Foundation) – Projektnummer 530143441 (project PuMMAVi).

\bibliography{bibliography}

@article{Krivanek2014,
   author = {Krivanek, O. L. and Lovejoy, T. C. and Dellby, N. and Aoki, T. and Carpenter, R. W. and Rez, P. and Soignard, E. and Zhu, J. and Batson, P. E. and Lagos, M. J. and Egerton, R. F. and Crozier, P. A.},
   title = {Vibrational spectroscopy in the electron microscope},
   journal = {Nature},
   volume = {514},
   number = {7521},
   pages = {209-12},
   ISSN = {1476-4687 (Electronic)
0028-0836 (Linking)},
   DOI = {10.1038/nature13870},
   url = {https://www.ncbi.nlm.nih.gov/pubmed/25297434},
   year = {2014},
   type = {Journal Article}
}

@article{Idrobo2018,
   author = {Idrobo, J. C. and Lupini, A. R. and Feng, T. and Unocic, R. R. and Walden, F. S. and Gardiner, D. S. and Lovejoy, T. C. and Dellby, N. and Pantelides, S. T. and Krivanek, O. L.},
   title = {Temperature Measurement by a Nanoscale Electron Probe Using Energy Gain and Loss Spectroscopy},
   journal = {Phys Rev Lett},
   volume = {120},
   number = {9},
   pages = {095901},
   ISSN = {1079-7114 (Electronic)
0031-9007 (Linking)},
   DOI = {10.1103/PhysRevLett.120.095901},
   url = {https://www.ncbi.nlm.nih.gov/pubmed/29547334},
   year = {2018},
   type = {Journal Article}
}

@article{Lagos2018,
   author = {Lagos, M. J. and Batson, P. E.},
   title = {Thermometry with Subnanometer Resolution in the Electron Microscope Using the Principle of Detailed Balancing},
   journal = {Nano Lett},
   volume = {18},
   number = {7},
   pages = {4556-4563},
   ISSN = {1530-6992 (Electronic)
1530-6984 (Linking)},
   DOI = {10.1021/acs.nanolett.8b01791},
   url = {https://www.ncbi.nlm.nih.gov/pubmed/29874456},
   year = {2018},
   type = {Journal Article}
}

@article{Castioni2025,
   author = {Castioni, F. and Auad, Y. and Blazit, J. D. and Li, X. and Woo, S. Y. and Watanabe, K. and Taniguchi, T. and Ho, C. H. and Stephan, O. and Kociak, M. and Tizei, L. H. G.},
   title = {Nanosecond Nanothermometry in an Electron Microscope},
   journal = {Nano Lett},
   volume = {25},
   number = {4},
   pages = {1601-1608},
   ISSN = {1530-6992 (Electronic)
1530-6984 (Linking)},
   DOI = {10.1021/acs.nanolett.4c05692},
   url = {https://www.ncbi.nlm.nih.gov/pubmed/39817671},
   year = {2025},
   type = {Journal Article}
}

@article{Liu2025,
   author = {Liu, F. and Mao, R. and Liu, Z. and Du, J. and Gao, P.},
   title = {Probing phonon transport dynamics across an interface by electron microscopy},
   journal = {Nature},
   volume = {642},
   number = {8069},
   pages = {941-946},
   ISSN = {1476-4687 (Electronic)
0028-0836 (Linking)},
   DOI = {10.1038/s41586-025-09108-6},
   url = {https://www.ncbi.nlm.nih.gov/pubmed/40500446},
   year = {2025},
   type = {Journal Article}
}

@article{McCauley2026,
   author = {McCauley, M. and Martis, J. and Krivanek, O. L. and Plotkin-Swing, B. and Mittelberger, A. and Wagner, T. and Celik, H. and Kornilov, G. and Zhao, M. and Hoflich, K. and Meffert, M. and Piazza, L. and Lovejoy, T. C. and Radtke, G. and Koch, C. T. and Haas, B.},
   title = {Platform and Framework for Time-Resolved Nanoscale Thermal Transport Measurements in STEM},
   journal = {Microsc Microanal},
   volume = {32},
   number = {3},
   ISSN = {1435-8115 (Electronic)
1431-9276 (Linking)},
   DOI = {10.1093/mam/ozag040},
   url = {https://www.ncbi.nlm.nih.gov/pubmed/42200812},
   year = {2026},
   type = {Journal Article}
}

@article{Haas2023,
   author = {Haas, B. and Boland, T. M. and Elsasser, C. and Singh, A. K. and March, K. and Barthel, J. and Koch, C. T. and Rez, P.},
   title = {Atomic-Resolution Mapping of Localized Phonon Modes at Grain Boundaries},
   journal = {Nano Lett},
   volume = {23},
   number = {13},
   pages = {5975-5980},
   ISSN = {1530-6992 (Electronic)
1530-6984 (Linking)},
   DOI = {10.1021/acs.nanolett.3c01089},
   url = {https://www.ncbi.nlm.nih.gov/pubmed/37341711},
   year = {2023},
   type = {Journal Article}
}

@article{Hage2020,
   author = {Hage, F. S. and Radtke, G. and Kepaptsoglou, D. M. and Lazzeri, M. and Ramasse, Q. M.},
   title = {Single-atom vibrational spectroscopy in the scanning transmission electron microscope},
   journal = {Science},
   volume = {367},
   number = {6482},
   pages = {1124-1127},
   ISSN = {1095-9203 (Electronic)
0036-8075 (Linking)},
   DOI = {10.1126/science.aba1136},
   url = {https://www.ncbi.nlm.nih.gov/pubmed/32139541},
   year = {2020},
   type = {Journal Article}
}

@article{Haas2023_MM,
   author = {Haas, Benedikt and Radtke, Guillaume and Nelson-Quillin, Amber M. and Lovejoy, Tracy C. and Dellby, Niklas and Krivanek, Ondrej L. and Hammud, Adnan and Schröder, Tim and Koch, Christoph T.},
   title = {Mapping Phonon Dispersion Surfaces at Nanometer Scale},
   journal = {Microscopy and Microanalysis},
   volume = {29},
   number = {Supplement 1},
   pages = {356-357},
   ISSN = {1431-9276
1435-8115},
   DOI = {10.1093/micmic/ozad067.166},
   year = {2023},
   type = {Journal Article}
}

@article{Lovejoy2020,
   author = {Lovejoy, Tracy and Plotkin-Swing, Benjamin and Dellby, Niklas and Meyer, Chris and Mittelberger, Andreas and Eljarrat, Alberto and Haas, Benedikt and Koch, Christoph and Krivanek, Ondrej},
   title = {Angle-Resolved Electron Energy Loss Spectroscopy},
   journal = {Microscopy and Microanalysis},
   volume = {26},
   number = {S2},
   pages = {964-965},
   ISSN = {1431-9276
1435-8115},
   DOI = {10.1017/s1431927620016505},
   year = {2020},
   type = {Journal Article}
}

@article{Hage2018,
   author = {Hage, F. S. and Nicholls, R. J. and Yates, J. R. and McCulloch, D. G. and Lovejoy, T. C. and Dellby, N. and Krivanek, O. L. and Refson, K. and Ramasse, Q. M.},
   title = {Nanoscale momentum-resolved vibrational spectroscopy},
   journal = {Sci Adv},
   volume = {4},
   number = {6},
   pages = {eaar7495},
   ISSN = {2375-2548 (Electronic)
2375-2548 (Linking)},
   DOI = {10.1126/sciadv.aar7495},
   url = {https://www.ncbi.nlm.nih.gov/pubmed/29951584},
   year = {2018},
   type = {Journal Article}
}

@article{Senga2019,
   author = {Senga, R. and Suenaga, K. and Barone, P. and Morishita, S. and Mauri, F. and Pichler, T.},
   title = {Position and momentum mapping of vibrations in graphene nanostructures},
   journal = {Nature},
   volume = {573},
   number = {7773},
   pages = {247-250},
   ISSN = {1476-4687 (Electronic)
0028-0836 (Linking)},
   DOI = {10.1038/s41586-019-1477-8},
   url = {https://www.ncbi.nlm.nih.gov/pubmed/31406319},
   year = {2019},
   type = {Journal Article}
}

@article{Yan2019,
   author = {Yan, X. and Liu, C. and Gadre, C. A. and Dai, S. and Gu, L. and Yu, K. and Aoki, T. and Wu, R. and Pan, X.},
   title = {Unexpected Strong Thermally Induced Phonon Energy Shift for Mapping Local Temperature},
   journal = {Nano Lett},
   volume = {19},
   number = {10},
   pages = {7494-7502},
   ISSN = {1530-6992 (Electronic)
1530-6984 (Linking)},
   DOI = {10.1021/acs.nanolett.9b03307},
   url = {https://www.ncbi.nlm.nih.gov/pubmed/31517496},
   year = {2019},
   type = {Journal Article}
}

@article{Montagnac2013,
   author = {Montagnac, G. and Caracas, R. and Bobocioiu, E. and Vittoz, F. and Reynard, B.},
   title = {Anharmonicity of graphite from UV Raman spectroscopy to 2700K},
   journal = {Carbon},
   volume = {54},
   pages = {68-75},
   ISSN = {00086223},
   DOI = {10.1016/j.carbon.2012.11.004},
   year = {2013},
   type = {Journal Article}
}

@article{Giura2012,
   author = {Giura, P. and Bonini, N. and Creff, G. and Brubach, J. B. and Roy, P. and Lazzeri, M.},
   title = {Temperature evolution of infrared- and Raman-active phonons in graphite},
   journal = {Physical Review B},
   volume = {86},
   number = {12},
   ISSN = {1098-0121
1550-235X},
   DOI = {10.1103/PhysRevB.86.121404},
   year = {2012},
   type = {Journal Article}
}

@article{Liu2019,
   author = {Liu, H. N. and Cong, X. and Lin, M. L. and Tan, P. H.},
   title = {The intrinsic temperature-dependent Raman spectra of graphite in the temperature range from 4K to 1000K},
   journal = {Carbon},
   volume = {152},
   pages = {451-458},
   ISSN = {00086223},
   DOI = {10.1016/j.carbon.2019.05.016},
   year = {2019},
   type = {Journal Article}
}

@article{Bonini2007,
   author = {Bonini, N. and Lazzeri, M. and Marzari, N. and Mauri, F.},
   title = {Phonon anharmonicities in graphite and graphene},
   journal = {Phys Rev Lett},
   volume = {99},
   number = {17},
   pages = {176802},
   ISSN = {0031-9007 (Print)
0031-9007 (Linking)},
   DOI = {10.1103/PhysRevLett.99.176802},
   url = {https://www.ncbi.nlm.nih.gov/pubmed/17995357},
   year = {2007},
   type = {Journal Article}
}

@article{Barantani2025,
   author = {Barantani, F. and Claude, R. and Iyikanat, F. and Madan, I. and Sapozhnik, A. A. and Puppin, M. and Weaver, B. and LaGrange, T. and de Abajo, F. J. G. and Carbone, F.},
   title = {Ultrafast momentum-resolved visualization of the interplay between phonon-mediated scattering and plasmons in graphite},
   journal = {Sci Adv},
   volume = {11},
   number = {14},
   pages = {eadu1001},
   ISSN = {2375-2548 (Electronic)
2375-2548 (Linking)},
   DOI = {10.1126/sciadv.adu1001},
   url = {https://www.ncbi.nlm.nih.gov/pubmed/40173224},
   year = {2025},
   type = {Journal Article}
}

@article{Koukaras2015,
   author = {Koukaras, E. N. and Kalosakas, G. and Galiotis, C. and Papagelis, K.},
   title = {Phonon properties of graphene derived from molecular dynamics simulations},
   journal = {Sci Rep},
   volume = {5},
   pages = {12923},
   ISSN = {2045-2322 (Electronic)
2045-2322 (Linking)},
   DOI = {10.1038/srep12923},
   url = {https://www.ncbi.nlm.nih.gov/pubmed/26316252},
   year = {2015},
   type = {Journal Article}
}

@article{Paulatto2013,
   author = {Paulatto, Lorenzo and Mauri, Francesco and Lazzeri, Michele},
   title = {Anharmonic properties from a generalized third-orderab initioapproach: Theory and applications to graphite and graphene},
   journal = {Physical Review B},
   volume = {87},
   number = {21},
   ISSN = {1098-0121
1550-235X},
   DOI = {10.1103/PhysRevB.87.214303},
   year = {2013},
   type = {Journal Article}
}

@article{Lovejoy2019,
   author = {Lovejoy, T. C. and Corbin, G. J. and Dellby, N. and Johnson, N. and Hoffman, M. V. and Mittelberger, A. and Krivanek, O. L.},
   title = {Progress in Ultra-High Energy Resolution EELS},
   journal = {Microscopy and Microanalysis},
   volume = {25},
   number = {S2},
   pages = {628-629},
   ISSN = {1431-9276
1435-8115},
   DOI = {10.1017/s1431927619003878},
   year = {2019},
   type = {Journal Article}
}

@article{Han2022,
   author = {Han, Z. and Yang, X. and Sullivan, S. E. and Feng, T. and Shi, L. and Li, W. and Ruan, X.},
   title = {Raman Linewidth Contributions from Four-Phonon and Electron-Phonon Interactions in Graphene},
   journal = {Phys Rev Lett},
   volume = {128},
   number = {4},
   pages = {045901},
   ISSN = {1079-7114 (Electronic)
0031-9007 (Linking)},
   DOI = {10.1103/PhysRevLett.128.045901},
   url = {https://www.ncbi.nlm.nih.gov/pubmed/35148139},
   year = {2022},
   type = {Journal Article}
}

@article{Haas2024,
   author = {Haas, Benedikt and Koch, Christoph T. and Rez, Peter},
   title = {Perspective on atomic-resolution vibrational electron energy-loss spectroscopy},
   journal = {Applied Physics Letters},
   volume = {125},
   number = {15},
   pages = {150502},
   ISSN = {0003-6951
1077-3118},
   DOI = {10.1063/5.0231688},
   year = {2024},
   type = {Journal Article}
}

@article{Qi2021,
   author = {Qi, R. and Shi, R. and Li, Y. and Sun, Y. and Wu, M. and Li, N. and Du, J. and Liu, K. and Chen, C. and Chen, J. and Wang, F. and Yu, D. and Wang, E. G. and Gao, P.},
   title = {Measuring phonon dispersion at an interface},
   journal = {Nature},
   volume = {599},
   number = {7885},
   pages = {399-403},
   ISSN = {1476-4687 (Electronic)
0028-0836 (Linking)},
   DOI = {10.1038/s41586-021-03971-9},
   url = {https://www.ncbi.nlm.nih.gov/pubmed/34789901},
   year = {2021},
   type = {Journal Article}
}

@article{Kepaptsoglou2025,
   author = {Kepaptsoglou, D. and Castellanos-Reyes, J. A. and Kerrigan, A. and Alves do Nascimento, J. and Zeiger, P. M. and El Hajraoui, K. and Idrobo, J. C. and Mendis, B. G. and Bergman, A. and Lazarov, V. K. and Rusz, J. and Ramasse, Q. M.},
   title = {Magnon spectroscopy in the electron microscope},
   journal = {Nature},
   volume = {644},
   number = {8075},
   pages = {83-88},
   ISSN = {1476-4687 (Electronic)
0028-0836 (Print)
0028-0836 (Linking)},
   DOI = {10.1038/s41586-025-09318-y},
   url = {https://www.ncbi.nlm.nih.gov/pubmed/40702193},
   year = {2025},
   type = {Journal Article}
}

@article{Jiang2024,
   author = {Jiang, H. and Wang, T. and Zhang, Z. and Liu, F. and Shi, R. and Sheng, B. and Sheng, S. and Ge, W. and Wang, P. and Shen, B. and Sun, B. and Gao, P. and Lindsay, L. and Wang, X.},
   title = {Atomic-scale visualization of defect-induced localized vibrations in GaN},
   journal = {Nat Commun},
   volume = {15},
   number = {1},
   pages = {9052},
   ISSN = {2041-1723 (Electronic)
2041-1723 (Linking)},
   DOI = {10.1038/s41467-024-53394-z},
   url = {https://www.ncbi.nlm.nih.gov/pubmed/39426978},
   year = {2024},
   type = {Journal Article}
}

@article{Gadre2022,
   author = {Gadre, C. A. and Yan, X. and Song, Q. and Li, J. and Gu, L. and Huyan, H. and Aoki, T. and Lee, S. W. and Chen, G. and Wu, R. and Pan, X.},
   title = {Nanoscale imaging of phonon dynamics by electron microscopy},
   journal = {Nature},
   volume = {606},
   number = {7913},
   pages = {292-297},
   ISSN = {1476-4687 (Electronic)
0028-0836 (Print)
0028-0836 (Linking)},
   DOI = {10.1038/s41586-022-04736-8},
   url = {https://www.ncbi.nlm.nih.gov/pubmed/35676428},
   year = {2022},
   type = {Journal Article}
}

@article{Hoglund2022,
   author = {Hoglund, E. R. and Bao, D. L. and O'Hara, A. and Makarem, S. and Piontkowski, Z. T. and Matson, J. R. and Yadav, A. K. and Haislmaier, R. C. and Engel-Herbert, R. and Ihlefeld, J. F. and Ravichandran, J. and Ramesh, R. and Caldwell, J. D. and Beechem, T. E. and Tomko, J. A. and Hachtel, J. A. and Pantelides, S. T. and Hopkins, P. E. and Howe, J. M.},
   title = {Emergent interface vibrational structure of oxide superlattices},
   journal = {Nature},
   volume = {601},
   number = {7894},
   pages = {556-561},
   ISSN = {1476-4687 (Electronic)
0028-0836 (Print)
0028-0836 (Linking)},
   DOI = {10.1038/s41586-021-04238-z},
   url = {https://www.ncbi.nlm.nih.gov/pubmed/35082421},
   year = {2022},
   type = {Journal Article}
}

@software{hyperspy,
  author       = {Francisco de la Peña and
                  Eric Prestat and
                  Jonas Lähnemann and
                  Vidar Tonaas Fauske and
                  Pierre Burdet and
                  Petras Jokubauskas and
                  Tom Furnival and
                  Carter Francis and
                  Magnus Nord and
                  Tomas Ostasevicius and
                  Katherine E. MacArthur and
                  Duncan N. Johnstone and
                  Mike Sarahan and
                  Thomas Aarholt and
                  Josh Taillon and
                  Vadim Migunov and
                  Alberto Eljarrat and
                  Jan Caron and
                  T. Nemoto and
                  Timothy Poon and
                  sivborg and
                  actions-user and
                  Nicolas Tappy and
                  Niels Cautaerts and
                  Suhas Somnath and
                  Tom Slater and
                  Michael Walls and
                  pietsjoh},
  title        = {hyperspy/hyperspy: v2.5.0},
  month        = oct,
  year         = 2026,
  publisher    = {Zenodo},
  version      = {v2.5.0},
  doi          = {10.5281/zenodo.23189274},
  url          = {https://doi.org/10.5281/zenodo.23189274},
}

@article{Yang2017,
   author = {Yang, J. A. and Parham, S. and Dessau, D. and Reznik, D.},
   title = {Novel Electron-Phonon Relaxation Pathway in Graphite Revealed by Time-Resolved Raman Scattering and Angle-Resolved Photoemission Spectroscopy},
   journal = {Sci Rep},
   volume = {7},
   pages = {40876},
   ISSN = {2045-2322 (Electronic)
2045-2322 (Linking)},
   DOI = {10.1038/srep40876},
   url = {https://www.ncbi.nlm.nih.gov/pubmed/28102368},
   year = {2017},
   type = {Journal Article}
}

\clearpage
\renewcommand{\thefigure}{S\arabic{figure}}
\setcounter{figure}{0}

\section*{Supplementary Information}

\subsection{S1 Additional time-resolved momentum-resolved measurements}

\begin{figure}[ht]
    \centering
    \includegraphics[width=0.6\linewidth]{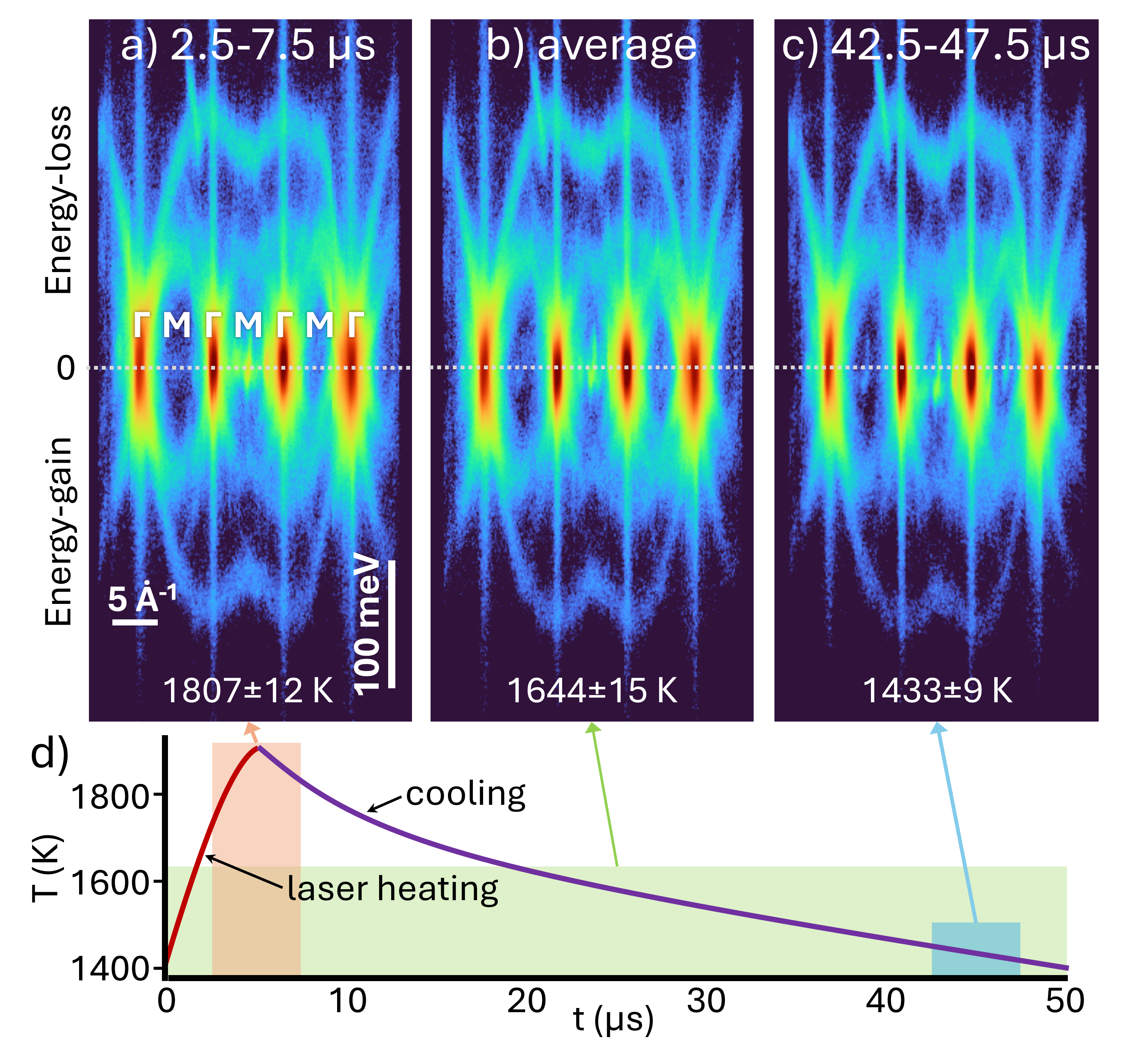}
    \caption{Time- and momentum-resolved vibrational EELS of graphite. a-c) Phonon dispersion at different times relative to a 5 $\mu$s laser pulse as indicated in d. Gating windows of 5 $\mu$s at delays of 2.5 and 42.5 $\mu$s relative to the beginning of the laser pulse were used for a and c respectively and b is an average over the entire cycle.}
    \label{fig:S2}
\end{figure}

Additional time-resolved momentum-resolved EELS spectra acquired on a different graphite flake supported by lacey carbon along the $\Gamma$M momentum axis are presented in Fig. \ref{fig:S2}. A 5 $\mu$s laser pulse with peak power density $\approx$124 MW/m$^2$ was used to heat the sample followed by a 45 $\mu$s cooling period as indicated in Fig. \ref{fig:S2}d. Fig. \ref{fig:S2}a was acquired 2.5 $\mu$ after the beginning of the pulse and Fig. \ref{fig:S2}c after 42.5 $\mu$s with the specimen temperature measured as 1807$\pm$12 K and 1433$\pm$9 K respectively. An average spectrum acquired with the same pulsed laser parameters without gated detection is shown in Fig. \ref{fig:S2}b with a measured temperature of 1644$\pm$15 K. 

\end{document}